\documentstyle[12pt,aasms4]{article}
\def\lapprox{\hbox{\lower .8ex\hbox{$\,\buildrel < \over\sim\,$}}}
\def\gapprox{\hbox{\lower .8ex\hbox{$\,\buildrel > \over\sim\,$}}}

\begin{document}

\title
{Type Ia supernova counts at high z: signatures of cosmological models and
progenitors}

\author
{P. Ruiz--Lapuente \altaffilmark{1,2}, and R. Canal \altaffilmark{1}}

\altaffiltext
{1}{ Department of Astronomy, University of Barcelona, Mart\'\i\ i Franqu\'es
1, E--08028 Barcelona, Spain. E--mail: pilar@mizar.am.ub.es,
ramon@mizar.am.ub.es}

\altaffiltext
{2}{Max--Planck--Institut f\"ur Astrophysik, Karl--Schwarzschild--Strasse 1,
D--85740 Garching, Federal Republic of Germany. E--mail:
pilar@MPA--Garching.MPG.DE}

\slugcomment{{\it Running title:} SNe Ia counts and cosmological models}

\begin{abstract}
Determination of the rates at which supernovae of Type Ia (SNe Ia) occur in
the early Universe can give signatures of the time spent by the binary
progenitor systems to reach explosion and of the geometry of the Universe.

Observations made within the Supernova Cosmology Project are already
providing the first numbers. Here it is shown that, for any assumed SNe Ia
progenitor, SNe Ia counts up to $m_{R}\simeq 23-26$ are useful tests of the
SNe Ia progenitor systems and cosmological tracers of a possible non--zero
value of the cosmological constant, $\Lambda$. The SNe Ia counts at high
redshifts compare differently with those at lower redshifts depending on the
cosmological model. Flat $\Omega_{\Lambda}$--dominated universes would show a
more significant increase of the SNe Ia counts at $z \sim 1$ than a flat,
$\Omega_{M} = 1$ universe. Here we consider three sorts of universes: a flat
universe with  $H_{0} = 65\ km\ s^{-1}\ Mpc^{-1}$, $\Omega_{M} = 1.0$,
$\Omega_{\Lambda} = 0.0$; an open universe with $H_{0} = 65\ km\ s^{-1}\
Mpc^{-1}$, $\Omega_{M} = 0.3$, $\Omega_{\Lambda} = 0.0$; and a flat,
$\Lambda$--dominated universe with $H_{0} = 65\ km\ s^{-1}\ Mpc^{-1}$,
$\Omega_{M} = 0.3$, $\Omega_{\Lambda} = 0.7$). On the other hand, the SNe Ia
counts from one class of binary progenitors (double degenerate systems)
should not increase steeply in the $z= 0$ to $z= 1$ range, contrary to what
should be seen for other binary progenitors. A measurement of the SNe Ia
counts up to $z \sim 1$ is within reach of ongoing SNe Ia searches at high
redshifts.
\end{abstract}

\keywords{cosmology: general --- supernovae: general}

\section{Introduction}

Supernovae are bright stellar explosions which can now be observed up to
redshifts $z \sim 1$. Several programmes (Perlmutter et al. 1997, 1998;
Schmidt et al. 1997; Garnavich et al. 1997) are presently devoted to discover
high--redshift supernovae for cosmological purposes. The goal is to determine
the geometry of the Universe through the effect that $\Omega_{M}$,
$\Omega_{\Lambda}$ have in the magnitude--redshift relationship of Type Ia
supernovae (SNe Ia). The same programmes provide rates of explosion up to
large redshifts through proper evaluation of control time and efficiency of
detection. Indeed, the Supernova Cosmology Project has already achieved the
first results after evaluating the data collected in its 1995 and 1996
discovery runs (Pain et al. 1996, 1997).
 
In this ${\it Letter}$ we outline the theoretical background for the
prediction of those number counts for different geometries of the Universe
and for various SNe Ia progenitors.

The pace at which stars formed in the past (the SFR) and the evolutionary
clock to explosion of those binary supernovae is reflected in the counts. The
evolutionary clock spans, in the case of SNe Ia ---as opposed to SNe II---,
an important fraction of the age of the Universe.

SNe Ia counts extending up to apparent red magnitudes $m_{R}\sim 23-25$ could
provide a good test of the SNe Ia binary progenitor systems. We also find
that SNe Ia counts should be sensitive to a $\Lambda$--dominancy in our
Universe through a larger increase at $z \sim 1$. Refinements in the
determination of the global SFR and continued SNe Ia searches at large
redshifts should furnish, in the near future, an accurate enough basis for
both tests.

\section{Modeling}

The global SFR recently derived for high redshift galaxies in the Hubble Deep
Field (HDF) by Madau et al. (1996) has been extended to lower redshifts
(Madau 1997) in accordance with the results of Lilly et al. (1995, 1996). The
variation of the SFR with $z$ is similar to that of the space density of
quasars (Shaver et al. 1996), peaking at $z\simeq 2$, the latter showing,
however, a steeper slope both prior and after maximum. The above $SFR(z)$ at
$z > 2$ has been obtained from the UV luminosity density along redshift.
Rowan--Robinson et al. (1997) have also derived a $SFR(z)$ from ISO infrared
observations of galaxies in the HDF. The general shape agrees well with
Madau's (1996) data but the inferred values are higher, maybe implying that
about 2/3 of star formation at high $z$ has taken place shrouded by dust. In
a different approach, Pei \& Fall (1995) have dealt with the determination of
the global star formation history by tracking the evolution of the global HI
contents of the Universe as measured from $Ly\alpha$ QSO absorption--line
systems. The results of Madau et al. (1996) agree with those from that very
different approach. On the other hand, Lilly et al. (1996) provide, from the
Canada--France--Hawaii Redshift Survey, an estimate of the comoving
luminosity density of the Universe over the redshift range $0 < z < 1$, which
can be interpreted in terms of a $SFR(z)$ which agrees with previous
estimates of a significant increase with $z$ up to $z\sim 1.5-2$. Prospects
to further explore the global SFR include  the number counts of SNe II
discovered at high redshift, which should trace the activity of star
formation in the Universe along $z$. Such numbers on SN II are obtainable in
current high--$z$ SNe searches.  Given the continuous improvement in the
determination of the global SFR, we base the present calculation of SNe Ia
number counts on the most recent empirical results. We evaluate the rate of
explosion of SNe Ia by convolving their time to explosion with the SFR. The
rate can then be calculated as:

$$r_{SNeIa}(t) = \int_{0}^{t} R(t - \tau)SFR(\tau) d\tau\eqno(1)$$

\noindent
where $R(t)$ is the SNe Ia rate after an instantaneous outburst, and $t$ and
$\tau$ are in the SN rest frame. Depending on the progenitor systems, in some
cases we would have  SNe Ia with a peak rate of explosion at 10$^{9}$ yr. In
other cases, there can be some SNe Ia exploding even only a few $10^{7}$ yr
after star formation already, thus becoming the brightest optical events at
$z  >> 1$. $SFR(\tau)$ is derived from the global $SFR(z)$ (Madau et al.
1996; Madau 1997). The adopted $SFR(z)$ (for $H_{0} = 50\ km\ s^{-1}\
Mpc^{-1}$, $\Omega_{M} = 1.0$, $\Omega_{\Lambda}$ = 0) is shown in the top
panel of Figure 1, along with the measured data points and their error bars.
Note that the data points from Connolly et al. (1997) for the $1\lapprox
z\lapprox 2$ range fall somewhat below whereas those from Rowan--Robinson et
al. (1997) for the $0.6\lapprox z\lapprox 1$ range are significantly above
the adopted curve. The $SFR(z)$ is transformed according to the geometries of
the model universes considered. Here we restrict this presentation to three
favored models of the Universe: {\it Model A}, a flat universe with $H_{0} =
65\ km\ s^{-1}$, $\Omega_{M} = 1$, $\Omega_{\Lambda} = 0$; {\it Model B}, an
open universe with $H_{0} = 65\ km\ s^{-1}\ Mpc^{-1}$, $\Omega_{M} = 0.3$,
$\Omega_{\Lambda} = 0$, and {\it Model C}, a flat universe with $H_{0} = 65\
km\ s^{-1}\ Mpc^{-1}$, $\Omega_{M} = 0.3$, $\Omega_{\Lambda} = 0.7$.

The $SFR(z)$  derived for each case is next transformed into $SFR(t)$ (in the
comoving frame). The derived global $SFR(t)$, when time--integrated over the
whole history of the corresponding universe (model $A$: $t_{0} = 10.0\ Gyr$;
model $B$: $t_{0} = 12.2\ Gyr$; model $C$: $t_{0} = 13.6\ Gyr$), produces the
observed stars today (Guzm\'an et al. 1997).

The global SNe Ia rates, $r_{SNe Ia}(t)$ ($yr^{-1}\ Mpc^{-3}$), for each
model universe and family of SNe Ia progenitor systems, are calculated in the
comoving frame according to (1), and integrated over comoving volume to
obtain the expected SNe Ia counts ($yr^{-1} sq. deg^{-1}$) as a function of
z. We integrate over $dV$:

$$dV = {d_{M}^{2}\over (1 + \Omega_{k}H_{0}^{2}d_{M}^{2})^{1/2}}\
d(d_{M})d\Omega\eqno(2)$$

\noindent
where $d_{M}$ is the proper motion distance and $\Omega_{k} = 1 - \Omega_{M}
- \Omega_{\Lambda}$ is the fractional contribution of the curvature to the
expansion (Carroll, Press, \& Turner 1992). $d_{M}$ is related to the
luminosity distance $d_{L}$ through $d_{L} = (1+z)\ d_{M}$ (Weinberg 1972).
$d_{L}$ is calculated as:

$$d_{L} = { (1 + z) \over H_{0}|\Omega_{k}|^{1/2}}\
sinn\left\{|\Omega_{k}|^{1/2}\int_{0}^{z_{1}}\left[(1+z)^{2} (1+\Omega_{M}z)
- z(2+z)\Omega_{\Lambda}\right]^{-1/2}dz\right\} \eqno(3)$$

\noindent
where $sinn$ stands for $sinh$ if $\Omega_{k} > 0$ and for $sin$ if
$\Omega_{k} < 0$ (both $sinn$ and the $\Omega_{k}$ terms disappear from (3)
if $\Omega_{k} = 0$, leaving only the integral times $(1 + z)/H_{0}$).

\noindent
The dependence on cosmological parameters of the comoving volume derivative
$dV/dz d\Omega$ differs from one model of Universe to another. A different
cosmological effect is the age--redshift (t(z)) relationship for different
model Universes, which changes the z at which the SNe Ia rates peak.

The results are tightly linked to the reliability of the global SFR. An
estimate of its current uncertainty can be obtained by looking at the data
points in the top panel of Figure 1. Increasingly accurate SFRs should
steadily improve the SNe Ia rate predictions. There is, as well, increasing
evidence of universality in the slope of the initial mass function (IMF) from
tests in different metallicity and age environments. Such a ``universal''
mass function would be well represented by the Salpeter (1955) power law with
$x = 0.86 \pm 0.23$. Both the overall trend to convergence in the estimate of
the global $SFR(z)$ and the almost constancy of the IMF favor the possibility
of deriving global values for the SNe Ia explosion rates.

\noindent{\it Stellar clock: binary progenitors}. All evolutionary models for
SNe Ia progenitors involve the accretion of mass by a C+O WD from a companion
in a close binary system. Current research on SNe Ia has discarded some
formerly proposed SNe Ia scenarios. The two classes of binary systems
considered here as the most likely systems giving SNe Ia are: (1) the merging
of two white dwarfs ---also called double degenerate scenario (DD)---, and
(2) the accretion by the WD of H  from a less evolved star in a close binary
system ---a single degenerate scenario (SD)---. We call this second case a
cataclysmic--like system (CLS). The time evolution of the SNe Ia rates after
star formation differs from one family of progenitor systems to another, and
there are variations within each family.  We have modeled the SNe Ia rates by
means of the same Monte Carlo code as in previous works (Ruiz--Lapuente,
Burkert, \& Canal 1995,1997;  Canal, Ruiz--Lapuente, \& Burkert 1996,
hereafter RCB95,97 and CRB96),  and adopting different  prescriptions for
each evolutionary path and for the initial binary parameters.  We have also
compared our predictions with those of other authors:  (1) The physical input
for the merging of two C+O white dwarfs (WDs) is modeled both as in Iben \&
Tutukov (1984)  and in RBC95. We allow  for a number of different physical
descriptions of the binary  evolution, and we try different values of the
common envelope  parameter $\alpha$ (a measure of the efficiency with which
orbital energy is used in envelope ejection).  (2) The physical modeling of
the explosion of a C+O WD when it reaches the Chandrasekhar mass by accretion
and burning of H from a main--sequence, subgiant, or giant  companion is
modeled as in the early work by Iben \& Tutukov (1984), as in CRB96,  and
finally including the most recently proposed variation of the binary
evolution by Hachisu, Kato, \& Nomoto (1996), who find a ``wind'' solution
for fast hydrogen accretion by a WD from a red--(sub)giant companion, but we
apply the same evolutionary constraints as Yungelson et al. (1996).
Particularly relevant to the modeling are the slope of the IMF, the
distribution of mass ratios $q$ of the secondary to the primary, and the
distribution of initial separations $A_{0}$ in the progenitor binary sytems,
together with the prescriptions for mass transfer.  By trying various
physical approaches adopted by different modelers and exploring different
values of the initial binary parameters (IMF, $q$, $A_{0}$ distributions) and
assumptions as to the mass transfer rates, we obtain the characteristic
behaviour of each stellar clock and the uncertainties in the absolute scale
of the rates.

In Figure 1, a comparison of $R(t)$ in (1) with those from other authors
(middle  panels: different IMF, $q$ and $A_{0}$ distributions; bottom panel:
different assumptions as to the allowed mass transfer rates) is made. From
such comparison we concluded in our previous work that the evolution of the
SNe Ia rates (rise, peak, decline) along cosmic time, for a given class of
systems,  have  broad common features shared by the predictions from
different authors (Iben \& Tutukov 1984; Tutukov \& Yungelson 1994; Yungelson
et al. 1996; our own modeling), which basically describe  the clock of the
corresponding systems. The Hachisu, Kato \& Nomoto (1996) solution, here
called CLS(W), would allow to have some relatively young SNe Ia in the CLS
scenario, and the SNe Ia rate would increase fast with redshift. On the other
hand, the DD scenario predicts a flatter increase of the rate from z $\simeq$
0 to z$\simeq$ 1, in all model universes. In all cases, the $R(t)$ shown here
would give, for our own Galaxy and the present time, SNe Ia rates which are
within the current error bars of the observational estimates.

Versions CLS and CLS(W) in the bottom panel of Figure 1 illustrate the range
of uncertainty in the time of start of the explosions in the SD scenarios.
For the DD progenitors we have chosen the most favorable physical
assumptions, those that enhance the numbers of SNe Ia at high $z$, in order
to best reproduce the first observational results. Any other choices would
give lower numbers. In fact, the observed numbers seem to be above the
predictions for this type of binary progenitor system (but one should stress
here the uncertainty in the global SFR). A clear feature of the DD scenario
is not only that the SNe Ia rates  do not increase fast towards higher
redshifts but also that they are lower than those predicted for the SD
progenitors. We should note here that we are assuming the universality of the
distributions of binary parameters determined for the solar neighbourhood
(Duquennoy \& Mayor 1991). We have explored, however, the effects that
plausible changes in the initial $q$ and $A_{0}$ distributions would have in
the outcome and found them to be only moderate (see below).

\noindent{\it Counts as a function of magnitude}. We compare the model counts
with the observations by deriving the dependence of the number counts on
$m_{R}$, the apparent magnitude of the SNe Ia in the $R$ photometric band. In
fact, in this transformation  the intrinsic dispersion in brightness of the
SNe Ia is not a key factor, since we are using the $N-mag$  relationship
measured in intervals of 0.2 to 0.5 mag for our tests (we could also use
$N-z$, thus avoiding the transformation to magnitude). The intrinsic
dispersion in magnitude is, in contrast, very important for cosmological
tests using $m(z)$.

We assume the SNe Ia to have an average absolute blue magnitude:

$$M_{B} = -18.52 + 5\ log(h/0.85)\eqno(4)$$

\noindent
where $h\equiv H_{0}/100$  (Perlmutter et al. 1997).

The distance modulus for each $z$ is calculated from the luminosity distance
$d_{L}$. The apparent $m_{R}$ at maximum as a function of $z$ is determined
from:

$$m_{R} = M_{B} + 5\ log (d_{L}(z)) - 5\  + K_{corr}\eqno(5)$$

\noindent
where $K_{corr}$ is taken from Kim, Goobar, \& Perlmutter (1996) (it includes
the full transformation from $B$ into $R$ magnitudes). We finally calculate
the variation of the SNe Ia rate ($yr^{-1}\ sq. deg^{-1}$ $\Delta mag^{-1}$)
with apparent red magnitude $m_{R}$ (we take $\Delta m_{R}=0.5\ mag$).

The results are displayed in the four panels of Figure 2. Shown in the Figure
are the data points currently available (Pain et al. 1996; Pain, private
communication). The dips in the curves are entirely due to the shape of the
$K$--corrections. The different slopes of the curves from one family of
progenitors to another mainly reflect how fast the SNe Ia rate declines after
reaching its peak value plus the time elapsed between star formation and the
peak (steeper decline and longer time delay for the CLS and CLS(W) models
than for the DD model: see the middle and bottom panels of Figure 1), and the
absolute values of the rates at low redshift ($m_{R} \simeq 20-21$) are
sensitive to the ages of the corresponding model universes ($t_{0}(A)<
t_{0}(B)< t_{0}(C)$). The slopes of the rates when comparing different models
of the Universe are sensitive to the contribution of a non--zero $\Lambda$.
The faster increase of the rates with magnitude (redshift) along the model
sequence A--C corresponds to the increasing comoving volume derivatives
($dV/dz d\Omega$) in the respective model universes. Such derivative is large
if $\Omega_{\Lambda}$ gives a significant contribution to $\Omega$. As a
check of the sensitivity of the results to the choices of the binary model
parameters, we have also calculated the $dN/dm_{R}$--$m_{R}$ relationship,
for the DD  progenitor, adopting  different $q$ and $A_{0}$ distributions
(dotted  line in the middle panel of Figure 1): the final curve is not very
sensitive to the various choices of the distributions explored in this work
(the model is shown  by the continuous line  in the four panels of Figure 2).
Taking at their face values the SFR and parameters for the SNe Ia progenitor
evolution adopted, and also the two data points, the results for both the CLS
and CLS(W) systems give better fits to the data than those for the DD
systems, model universe A being most favored and model C giving the worst fit
for any kind of system. But those conclusions are preliminary and we must
await the reduction of the uncertainties in the observed counts and in the
global SFR for the intermediate--mass stars leading to SNe Ia. As we see from
Figure 2, at present the uncertainty in the SNe Ia counts at $m_{R}\simeq 22$
is of the order of a factor of 2. That is the same as the difference between
the prediction for the DD model and the lowest one for the SD model, and also
similar to the range covered by the two extreme predictions (CLS and CLS(W))
for the SD model. Any increase in the SFR would almost homologously shift
upwards all the count predictions. At higher magnitudes, the range of
predictions for the SD models becomes narrower while the differences with the
DD model increase. Thus, if we knew both the SNe Ia counts at $m_{R}\simeq
23$ and the global SFR to better than a factor 1.5, we could already
discriminate between the SD and the DD models. The sensitivity to the
cosmological parameters $\Omega_{M}$ and $\Omega_{\Lambda}$ is still low at
those magnitudes. If we had determined, for instance, the DD model to be the
right one, then the SNe Ia counts at $m_{R} = 24.5$ (corresponding to
$z\simeq 1$) in the C universe ($\Omega_{M} = 0.3$, $\Omega_{\Lambda} = 0.7$)
should be twice those in the A universe ($\Omega_{M} = 1$, $\Omega_{\Lambda}
= 0$), the factor of increase from $m_{R} = 22$ to $m_{R} = 24.5$ being also
twice as large in case C as compared with case A (both contrasts would be
sharper if the SD model were the right one). The high--$z$ SN searches to
measure $\Omega_{M}$, $\Omega_{\Lambda}$ from the variation of the apparent
magnitudes of SNe Ia are now almost reaching $z = 1$ (Perlmutter et al. 1998;
Garnavich et al. 1997). From samples large enough to derive the SNe Ia counts
at $z\simeq 1$, the constraints on both parameters should be tighter than
those obtainable from the counts alone, but the latter could provide a
supplementary test, since it is based on a different approach: the use of a
volume effect instead of a redshift--magnitude effect. Counts extending to
even higher magnitudes would more clearly reveal the geometry of the
Universe. Suggested SN searches up to infrared magnitudes $K\sim 26-27$
(Miralda--Escud\'e \& Rees 1997) would extend well beyond that point.

\section{Conclusions and future prospects}

The theoretical results presented here are intended to show the potential of
the comparison between model predictions and the results of undergoing and
future SNe Ia searches. The slopes of the $dN/dm_{R}$--$m_{R}$ curves are
mainly sensitive to the general characteristics of each SNe Ia model while
the absolute values depend more on model parameters and on the SFR. Hence the
interest of data covering a broader $m_{R}$ (or $z$) range than those
currently available. On the other hand, the differences in the predictions
from any SNe Ia progenitor assumed, for $\Omega_{\Lambda}$--dominated
universes as compared with matter--dominated and open universes, become
significant at large enough values of $m_{R}$. In a Universe dominated by
$\Omega_{\Lambda}$  the number counts of SNe Ia show a larger increase at $z
= 1$ due to the large volume encompassed at that redshift. Useful information
would be obtained by the measurement of the evolution of the SNe Ia counts up
to $z = 1$ and beyond (possibly in the K--band for higher redshifts).

The first data obtained by Pain et al. (1996) gave a value of
$34.4^{+23.9}_{-16.2}$ SNe Ia $yr^{-1}\ sq. deg.^{-1}$ in a magnitude range
of 21.3 $<$ R $<$ 22.3.  These first results were obtained from three SNe
discovered at redshifts 0.374, 0.420, and 0.354 (SN 1994H, SN 1994al \& SN
1994F) and  the small number statistics dominates this very first
measurement. A larger bulk of data, already obtained, will now reduce the
statistical uncertainties, and prospects to extend the observations up to
higher $z$ are on the way. It is thus too soon to extract firm conclusions
from the data. Our purpose here is to propose a new useful test to be
completed in the near future.

\clearpage

\begin{figure}[hbtp]
\centerline{\epsfysize15cm\epsfbox{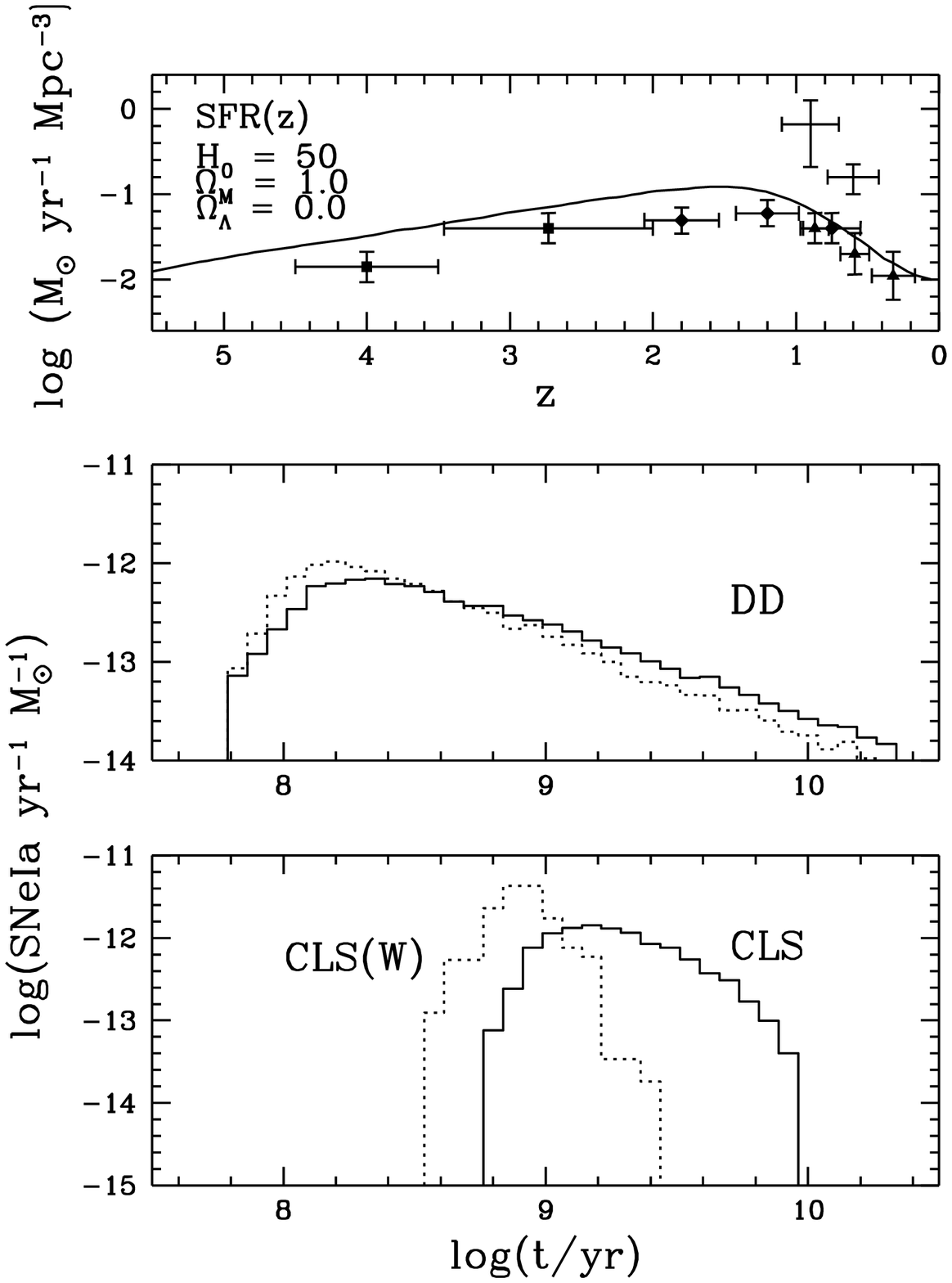}}
\nopagebreak[4]
\figcaption{{\it Top panel}: the adopted global $SFR(z)$, from Madau (1997),
together with different observational determinations: $triangles$ from Lilly
et al. (1995, 1996), $diamonds$ from Connolly et al. (1997), $squares$ from
Madau et al. (1996), $crosses$ from Rowan--Robinson et al. (1997). {\it
Middle panel}: the SNe Ia rates after an instantaneous outburst of star
formation, for the double--degenerate (DD) model. The continuous line is our
own result and the dotted line that of Tutukov \& Yungelson (1994), who adopt
a different IMF and also different $q$ and $A_{0}$ distributions. {\it Bottom
panel}: same, for the cataclysmic--like system (CLS) model. Continuous line
is our own result and dotted line the ``wind'' solution of Hachisu, Kato, \&
Nomoto (1996), with evolutionary prescriptions from Yungelson et al. (1996)
and ourselves (see text).
\label{fig1}}
\end{figure}

\begin{figure}[hbtp]
\centerline{\epsfysize15cm\epsfbox{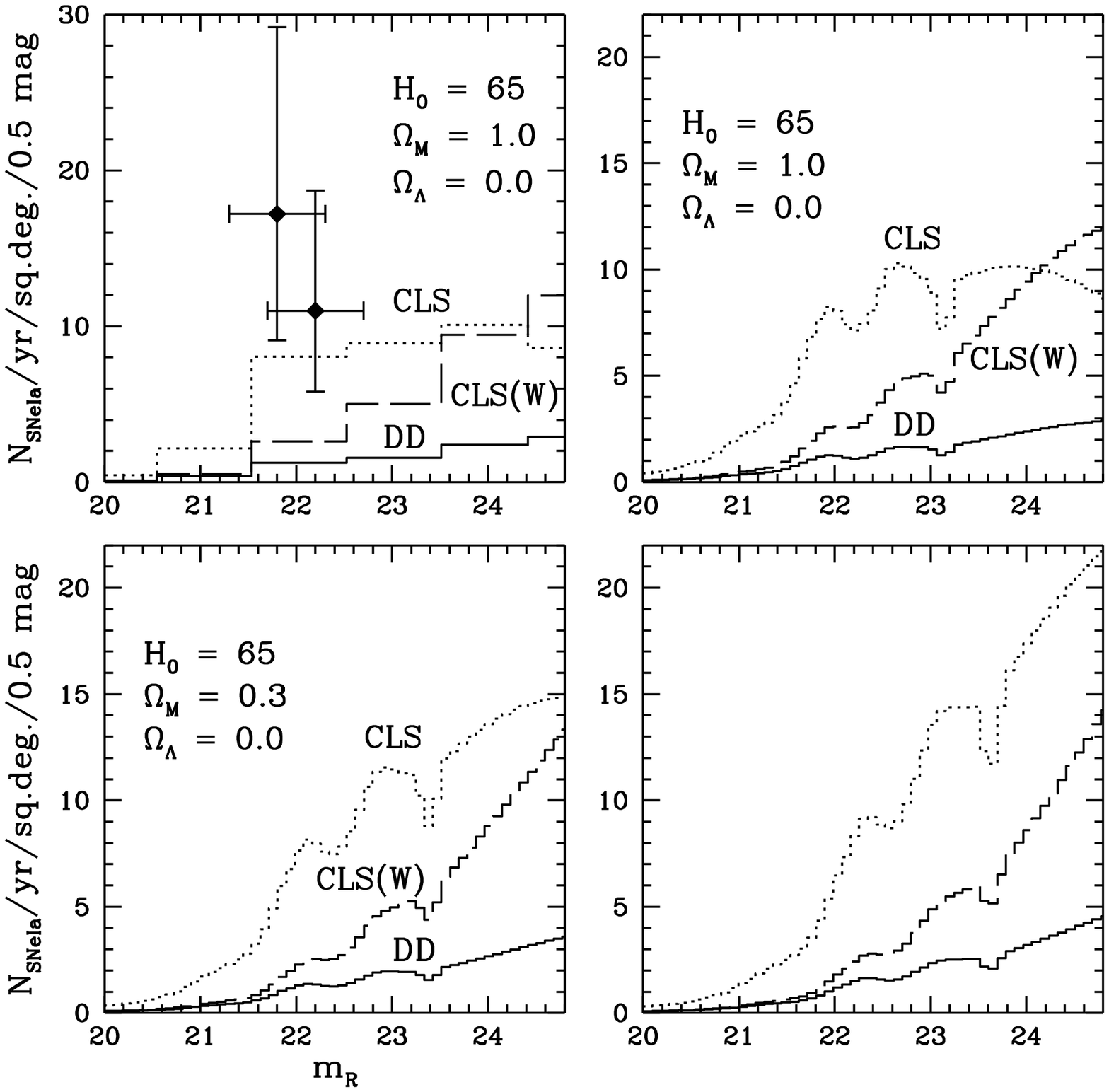}}
\nopagebreak[4]
\figcaption{Variation of the SNe Ia rates with limiting apparent red
magnitude $m_{R}$, for model universes A (top left panel), binned to
intervals of 1 mag to compare with the results of Pain et al (1996, 1997);
model universe A but without the binning (top right panel); model universe B
(bottom left panel) and C (bottom right panel), for the different SNe Ia
progenitor systems considered. Continuous lines correspond to the
double--degenerate (DD) systems, dotted lines to the cataclysmic--like
systems (CLS), and long--dashed lines to the CLS systems also, for the
``wind'' solution of Hachisu, Kato, \& Nomoto (1996) (see text).
\label{fig2}}
\end{figure}


\begin{thebibliography}{}
\bibitem{}
Canal, R., Ruiz--Lapuente, P., \& Burkert, A. 1996, ApJ, 456, L101 (CRB96)
\bibitem{}
Carrol, S.M., Press, W.H., \& Turner, E.L. 1992, ARA\&A, 30, 499
\bibitem{}
Connolly, A.J, \& Szalay, A.S., Dickinson, M., SubbaRao, M.U. \& Brunner, R.J.
1997, ApJ, 486, L1
\bibitem{}
Duquennoy, A., \& Mayor, M. 1991, A\&A, 248, 485
\bibitem{}
Garnavich, P.M., et al. 1997, ApJ, submitted, and preprint astro--ph/9719123
\bibitem{}
Guzm\'an, R., et al. 1997, ApJ, in press, and preprint astro-ph/9704001
\bibitem{}
Iben, I.,Jr., \& Tutukov, A.V. 1984, ApJS, 54, 335
\bibitem{}
Hachisu, I., Kato, M., \& Nomoto, K. 1996, ApJ, 470, L97
\bibitem{}
Kim, A., Goobar, A., \& Perlmutter, S. 1996, PASP, 108, 190
\bibitem{}
Lilly, S.J., Tresse, L., Hammer, F., Crampton, D., \& Le F\`evre, O. 1995,
ApJ, 455, 108
\bibitem{}
Lilly, S.J., Le F\`evre, O., Hammer, F., \& Crampton, D. 1996, ApJ, 460, L1
\bibitem{}
Madau, P. 1997, PASP Conf. Ser., in press, and preprint
astro--ph/9707141
\bibitem{}
Madau, P., et al. 1996, MNRAS, 283, 1388
\bibitem{}
Miralda--Escud\'e, J., \& Rees, M.J. 1997, ApJ, 478, L57
\bibitem{}
Pain, R., et al. 1996, ApJ, 473, 356
\bibitem{}
--------------------. 1997, in preparation
\bibitem{}
Pei, Y.C., \& Fall, M.S. 1995, ApJ, 454, 69
\bibitem{}
Perlmutter, S., et al. 1997, ApJ, 483, 565
\bibitem{}
Perlmutter, S., et al. 1998, Nature, in press
\bibitem{}
Rowan--Robinson, M., et al. 1997, MNRAS, 289, 490
\bibitem{}
Ruiz--Lapuente, P., Burkert, A., \& Canal, R. 1995, ApJ, 447, L69 (RBC95)
\bibitem{}
Ruiz--Lapuente, P., Canal, R., \& Burkert, A. 1997, in Thermonuclear
Supernovae, ed. P. Ruiz--Lapuente, R. Canal, \& J. Isern (Dordrecht: Kluwer),
205 (RCB97)
\bibitem{}
Salpeter, E.E. 1955, ApJ, 121, 161
\bibitem{}
Schmidt, B., et al. 1996, Bull. AAS, 189, 108--05
\bibitem{}
Shaver, P.A., Wall, J.V., Kellermann, K.I., Jackson, C.A., \& Hawkins, M.R.S.
1996, Nature, 384, 439
\bibitem{}
Tutukov, A.V., \& Yungelson, R.L. 1994, MNRAS, 268, 871
\bibitem{}
Weinberg, S. 1972, Gravitation and Cosmology (New York: Wiley)
\bibitem{}
Williams, R., et al. 1996, AJ, 112, 1335
\bibitem{}
Yungelson, L., Livio, M., Truran, J.W., Tutukov, A., \& Fedorova, A. 1996,
ApJ, 466, 890
\end{thebibliography}
\end{document}